\documentclass[sigconf, nonacm]{acmart}

\usepackage{enumitem}
\usepackage[linesnumbered, ruled, vlined,resetcount]{algorithm2e} 
\usepackage{graphics}
\usepackage{balance}
\usepackage{graphicx}
\usepackage{subfig}
\usepackage{amsthm}
\usepackage{bm}
\usepackage{algpseudocode}
\usepackage{listings}
\usepackage{{makecell}}
\usepackage[ruled, vlined, linesnumbered]{algorithm2e}
\usepackage{wrapfig}

\newcommand\vldbdoi{XX.XX/XXX.XX}
\newcommand\vldbpages{XXX-XXX}
\newcommand\vldbvolume{14}
\newcommand\vldbissue{1}
\newcommand\vldbyear{2020}
\newcommand\vldbauthors{\authors}
\newcommand\vldbtitle{\shorttitle}
\newcommand\vldbavailabilityurl{}
\newcommand\vldbpagestyle{plain}

\newcommand{\eat}[1]{}
\newcommand{\ie}{\emph{i.e.,}\xspace}
\newcommand{\eg}{\emph{e.g.,}\xspace}

\newcommand{\etc}{\emph{etc.}\xspace}

\newcommand{\Paragraph}[1]{\smallskip\noindent{\bf #1.}}
\newcommand{\taas}{\texttt{TaaS}\xspace}

\begin{document}
\fancyhead{}

\title{Towards Transaction as a Service}

\eat{
\author{author}
\affiliation{%
  \institution{institution}
  \city{city}
  \country{country}
}
\email{donald@swa.edu}
}

\author{Yanfeng Zhang$^{\dagger}$, Weixing Zhou$^{\dagger}$, Yang Ren$^{\S}$, Sihao Li$^{\S}$, Guoliang Li$^{\ddagger}$, Ge Yu$^{\dagger}$}
\affiliation{
 \institution{$\dagger$ Northeastern University, China \hspace{3ex} $\S$ Huawei Technology Co., Ltd \hspace{3ex} $\ddagger$ Tsinghua University, China}
  \city{}
  \country{}}
\email{}

\begin{abstract}
This paper argues for decoupling transaction processing from existing two-layer cloud-native databases and making transaction processing as an independent service. By building a transaction as a service (TaaS) layer, the transaction processing can be independently scaled for high resource utilization and can be independently upgraded for development agility. Accordingly, we architect an execution-transaction-storage three-layer cloud-native database. By connecting to TaaS, 1) the AP engines can be empowered with ACID TP capability, 2) multiple standalone TP engine instances can be incorporated to support multi-master distributed TP for horizontal scalability, 3) multiple execution engines with different data models can be integrated to support multi-model transactions, and 4) high performance TP is achieved through extensive TaaS optimizations and consistent evolution. Cloud-native databases deserve better architecture: we believe that TaaS provides a path forward to better cloud-native databases.
\end{abstract}

\maketitle

\eat{
\pagestyle{\vldbpagestyle}
\begingroup\small\noindent\raggedright\textbf{PVLDB Reference Format:}\\
\vldbauthors. \vldbtitle. PVLDB, \vldbvolume(\vldbissue): \vldbpages, \vldbyear.\\
\href{https://doi.org/\vldbdoi}{doi:\vldbdoi}
\endgroup
\begingroup
\renewcommand\thefootnote{}\footnote{\noindent
This work is licensed under the Creative Commons BY-NC-ND 4.0 International License. Visit \url{https://creativecommons.org/licenses/by-nc-nd/4.0/} to view a copy of this license. For any use beyond those covered by this license, obtain permission by emailing \href{mailto:info@vldb.org}{info@vldb.org}. Copyright is held by the owner/author(s). Publication rights licensed to the VLDB Endowment. \\
\raggedright Proceedings of the VLDB Endowment, Vol. \vldbvolume, No. \vldbissue\ %
ISSN 2150-8097. \\
\href{https://doi.org/\vldbdoi}{doi:\vldbdoi} \\
}\addtocounter{footnote}{-1}\endgroup

\ifdefempty{\vldbavailabilityurl}{}{
\vspace{.3cm}
\begingroup\small\noindent\raggedright\textbf{PVLDB Availability Availability:}\\
The source code of this research paper has been made publicly available at \url{\vldbavailabilityurl}.
\endgroup
}
}

\section{Introduction}
\label{sec:intro}

Database systems have evolved to be with a cloud-native architecture, \ie disaggregation
of compute and storage architecture \cite{verbitski2017amazon, verbitski2018amazon, li2019cloud, li2022cloud}, which decouples the storage from the compute nodes, then connects the compute nodes
to shared storage through a high-speed network. Compared with traditional databases where the compute and storage are bundled together, the two-layer architecture (execution layer and storage layer) allows the compute and storage resources to be scaled independently, thereby bringing more elasticity. A set of works are proposed to improve the cloud-native database performance, \eg transaction atomicity \cite{10.14778/3565816.3565837}, computation pushdown \cite{9101371}, local caching \cite{10.14778/3476249.3476265}, shared memory pool \cite{10.14778/3467861.3467877}, \etc As more and more enterprises move their applications to the cloud, cloud-native database systems start to gain wide support and popularity.

The principle of cloud-native architecture is \textit{decoupling}. Cloud can provide unlimited decoupled resources (including compute, memory, storage, \etc) and also provide the elasticity of different resources, allowing growing or shrinking the resource capacity to adapt to the changing demands. To fully utilize the decoupled resources, systems should be decoupled into independent function modules as much as possible, so that each single function module that mainly relies on a specific resource can be scaled independently to meet varying demands, making good use of different resources and achieving elasticity. On the other hand, these function modules can be designed as independent services \cite{alshuqayran2016systematic, cerny2018contextual}, so that each function module can be reused more easily and can be upgraded independently, bringing more agility. Transaction processing (TP) is a key function of a database, which aims to provide atomicity, consistency, isolation, and durability (ACID) features. Most existing cloud-native databases couple TP either with storage layer \cite{huang2020tidb} or with execution layer \cite{verbitski2017amazon, verbitski2018amazon}. 

\begin{figure}[t!]
\vspace{0.1in}
  \centerline{\includegraphics[width=3.2in]{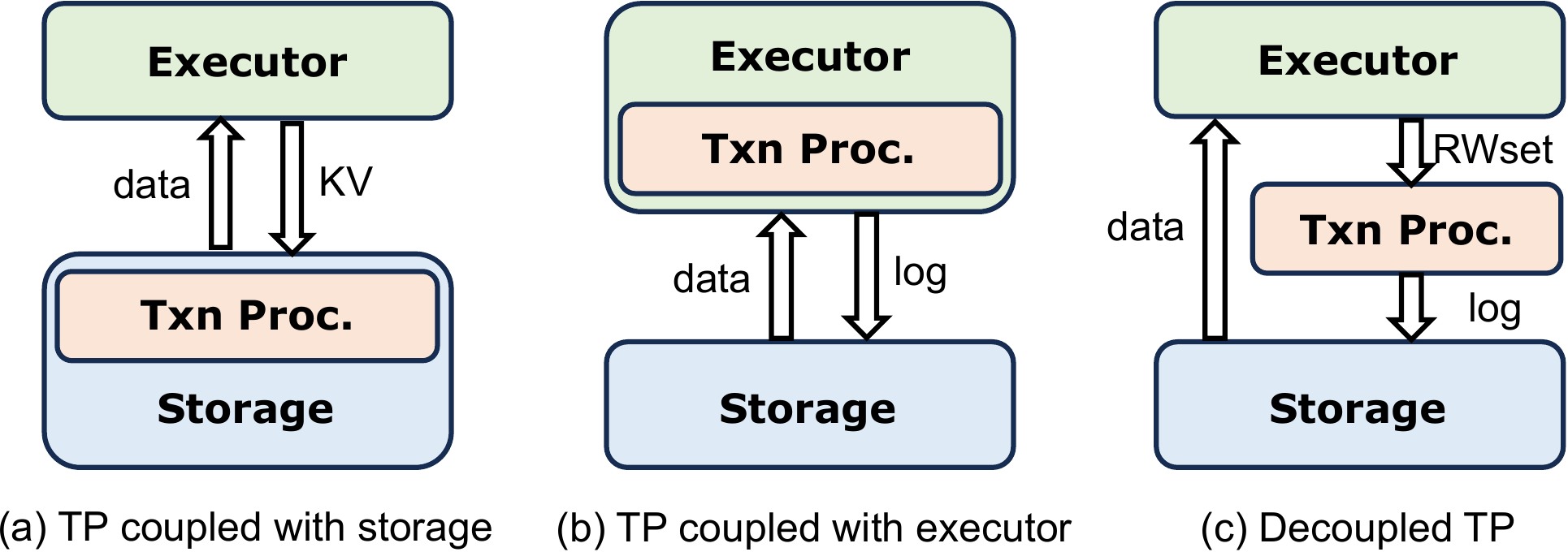}}
  \centering
  \vspace{-0.1in}
  \caption{Transaction processing module in disaggregated database architectures.}
  \label{fig:intro}
  \vspace{-0.1in}
\end{figure}

\Paragraph{Coupling TP with Storage Layer} Some two-layer databases couple TP with the storage layer as shown in Figure \ref{fig:intro}a. For example, TiDB \cite{huang2020tidb} adopts a distributed transactional key-value (KV) store TiKV \cite{tikv} as the storage layer, where the TiDB server is the stateless SQL layer that receives SQL requests, performs SQL parsing and optimization, and generates a distributed execution plan. TiKV is responsible for storing shareded data and supports distributed transactions. Such a design provides an independent KV store with ACID properties and a decoupled SQL execution engine, which is ideal for reuse and independent evolution. However, coupling TP with storage has two limitations. First, storage servers are usually configured with high-volume SSDs/disks but relatively low compute resources, while TP requires high parallelism. This causes a contradiction and will impact cost efficiency, violating the purpose of cloud-native design (\ie decoupled functions to make good use of disaggregated resources). Second, the storage is not commonly elastically scaled, while the TP should be elastically scaled according to varying loads. Bundling storage and TP causes another contradiction and restricts independent scaling property.

\Paragraph{Coupling TP with Execution Layer} Some other two-layer databases couple TP with the execution layer as shown in Figure \ref{fig:intro}b. For example, Amazon Aurora \cite{verbitski2017amazon, verbitski2018amazon} leverages MySQL \cite{mysql} or PostgreSQL \cite{pg} as the SQL execution instance, which handles TP in the execution layer. The redo logging, durable storage, crash recovery, and backup/restore are off-loaded to the storage service. By only writing redo log records to storage, network IOPS can be greatly reduced. MySQL and PostgreSQL are well-known open-source SQL relational databases. The execution engine should be specifically optimized to generate execution plans for different data models or different hardwares. For example, graph databases (\eg Nebula Graph \cite{nebula} and Neo4j \cite{neo4j}) are designed for graph query optimizations (\eg subgraph matching, shortest path, reachability, and Pagerank), vector databases (\eg Milvus \cite{milvus}) are designed for similarity search, and GPU databases (\eg GDB \cite{he2009relational}, MapD \cite{mostak2013overview}, GPUDB \cite{yuan2013yin}) are optimized for many-core parallelization \cite{shanbhag2020study}. However, the cores of TP engines are similar, \ie handling read-write or write-write conflicts of concurrent transactions. Bundling TP and the execution layer together would incur redevelopment costs for resolving transaction conflicts.

Considering the principle of cloud-native designs (\ie continuously decoupling functionality), it is desirable to decouple TP from the database architecture and make it work as an independent transaction service that allows different execution engines with various data models to connect. The transaction service can be independently scaled for high resource utilization and can be independently upgraded and evolved for development agility. 


In this paper, we propose a vision of \textit{transaction as a service} (\taas), which forms a new three-tier database architecture. As shown in Figure \ref{fig:intro}c, the execution layer executes the transaction queries and generates the readset and writeset which are posted to the \taas layer. Multiple execution engine instances might concurrently post readset/writeset to \taas, where the concurrency conflicts are resolved based on a certain \textit{pluggable} concurrency control algorithm. The results of the successfully committed transactions are durable as \textit{logs} in the \taas layer, and the commit/abort notifications are returned to the execution layer immediately. For high performance, the logs are asynchronously flushed to the storage layer, where the logs are replayed to update the data storage. The read consistency issues are tackled optimistically in the \taas layer according to users' consistency requirements.

Despite that introducing a new transaction layer leads to additional network I/O costs, the three-tier layer design brings more advantages. First, by connecting existing NoSQL databases to \taas, the NoSQL databases can be empowered with ACID TP capability. We have empowered HBase \cite{hbase}, LevelDB \cite{leveldb}, and NebulaGraph \cite{nebula} with ACID TP support, repectively ($\S$\ref{sec:adv:ap}). Second, by connecting multiple existing standalone TP engine instances to \taas, a multi-master distributed TP can be realized to improve the TP's horizontal scalability. We have achieved distributed TP by relying on multiple openGauss \cite{avni2020industrial} (which only supports single-machine TP) instances ($\S$\ref{sec:adv:disttp}). Third, by connecting multiple execution engines with different data models to \taas, multi-model transactions are supported. We decompose users' KV-SQL-Graph transactions into KV, SQL, or graph subtransactions and distribute them to TiKV \cite{tikv}, openGauss \cite{opengauss}, and NebulaGraph \cite{nebula} execution engine instances accordingly. The updates (in the format of writesets) generated by different model engines are merged with conflicts resolving in the \taas layer, thereby achieving multi-model transactions ($\S$\ref{sec:adv:multimodel}). Fourth, the \taas layer can be optimized and upgraded independently for high performance. Our preliminary results show that \taas achieves 4.7-8.5 times higher throughput and 4.5-5.2 times lower latency than the state-of-the-art disaggregated databases TiDB \cite{huang2020tidb} and FoundationDB \cite{zhou2021foundationdb} under YCSB and TPC-C benchmarks ($\S$\ref{sec:adv:perf}). 

In the rest of this paper, we present our three-tier database architecture design ($\S$\ref{sec:arch}) before demonstrating the advantages of making transaction processing as a service ($\S$\ref{sec:advantages}). 
We finally summarize the challenges and research opportunities that \taas brings ($\S$\ref{sec:challenges}).

\section{Transaction Service}
\label{sec:arch}

This section presents \taas and the new cloud-native database architecture based on \taas. 

\begin{figure}[t!]
  \centerline{\includegraphics[width=3.4in]{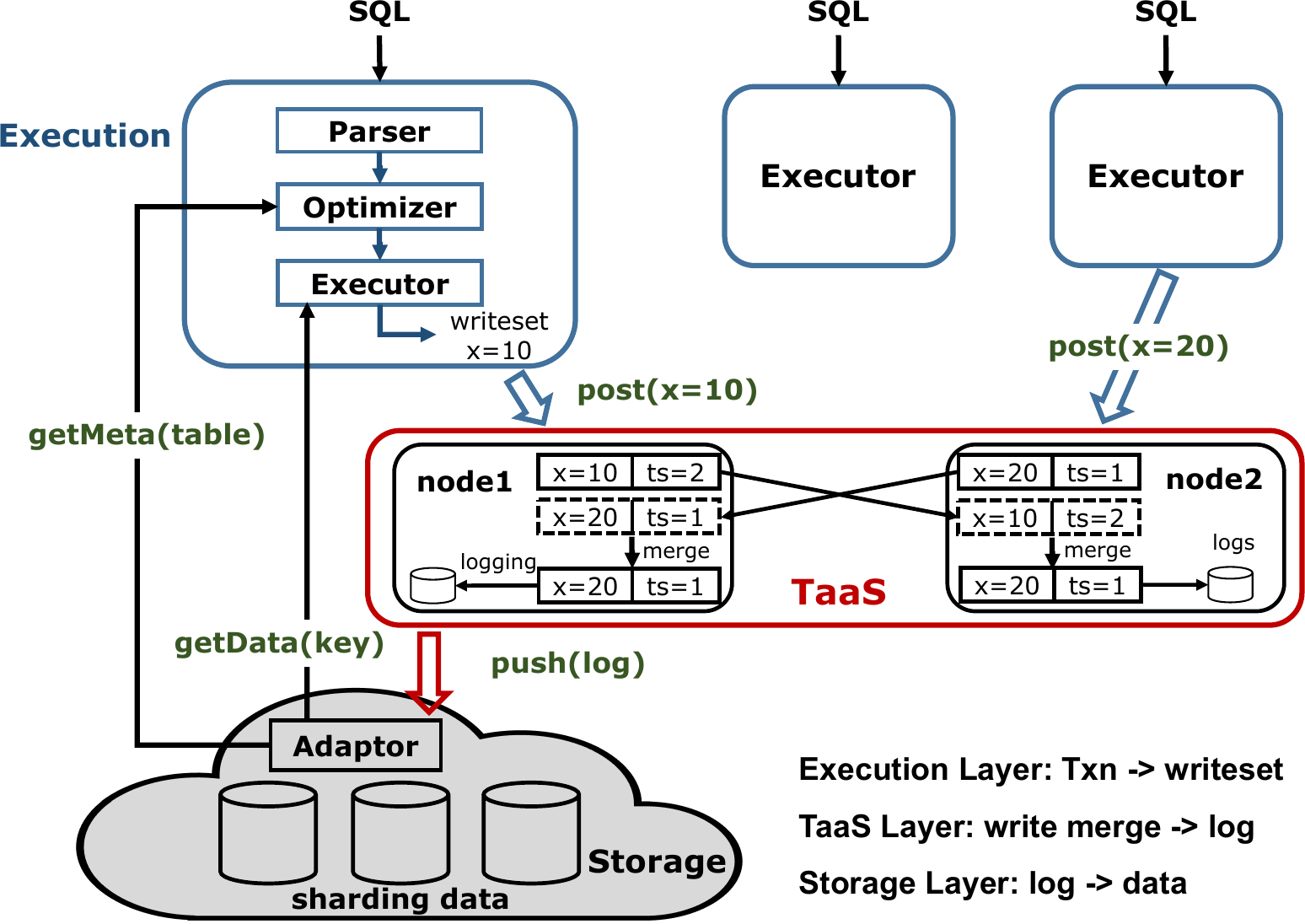}}
  \centering
  \vspace{-0.1in}
  \caption{TaaS architecture and workflow.}
  \label{fig:arch}
  \vspace{-0.1in}
\end{figure}

\subsection{TaaS Architecture and Workflow}
\label{sec:arch:overview}

As shown in Figure \ref{fig:arch}, after decoupling transaction processing functionality, an execution-transaction-storage three-layer database architecture is constructed. The execution layer consists of multiple \textit{stateless} execution engine instances, each of which accepts users' transaction requests in the format of SQL or other query languages. The transaction (or \taas) layer consists of multiple \taas nodes, accepting multiple concurrent updates from different execution engines and performing concurrency conflict resolution to output logs that contain the successful updates. The storage layer stores sharding data tables and metadata, and each storage node receives logs and updates tables according to these logs.

\Paragraph{Execution Layer (From Txn to Writeset)}
The design of the execution layer is similar to TiDB Server \cite{huang2020tidb}. After parsing the user's transaction request, each execution engine instance accesses the metadata (\eg table schema and data distribution) from the storage layer by invoking the \texttt{getMeta()} interface, which is used to generate a physical execution plan. The storage layer should prepare the metadata required for different execution engines. The executor \textit{optimistically} executes the plan, reads data from the storage layer by invoking the \texttt{getData()} interface, and outputs the readset and writeset of each transaction. Once the user commits the transaction, the cached readset and writeset are posted to the \taas layer through the \texttt{post()} interface (the readset is not drawn in Figure \ref{fig:arch} for simplicity). Slight modifications are required for the execution engines. It is noticeable that distributed execution engines are also allowed, where multiple sub-transactions sent from different machines are bundled into a transaction identified by an additional mark.

\Paragraph{Transaction Layer (From Writesets to Log)} 
\taas layer contains multiple \taas nodes. To avoid a single node bottleneck, any \taas node can accept readsets and writesets from the execution layer, which forms a multi-master architecture. Since only the read and write operations are transferred to the \taas layer, the concurrency control problem of transaction processing becomes a read-write or write-write conflict resolution problem. There are many design choices for conflict resolution, such as pessimistic two-phase locking \cite{yu2018sundial, lu2018star}, optimistic mechanisms \cite{tu2013speedy, lu2021epoch, zhou2023geogauss}, and deterministic methods \cite{thomson2012calvin, lu2020aria, ren2019slog}. The conflict resolution results that indicate the transaction commit or abort are logged, which denotes the updated data. Simultaneously, the transaction commit or abort notifications are returned to the execution layer and users for low latency. The detailed design of \taas will be discussed in Section \ref{sec:arch:taas}. The logs are asynchronously pushed to the storage layer through a \texttt{push()} interface. The data consistency, durability, and fault recovery issues will be discussed in Section \ref{sec:arch:impl}.

\Paragraph{Storage Layer (From Log to Data)}
Since the storage structure can be dramatically different, such as columnar storage \cite{abadi2008column}, row-column hybrid storage \cite{he2011rcfile}, graph-native storage \cite{deutsch2019tigergraph}, and vector store) \cite{wang2021milvus}, a \textit{storage adaptor} needs to be implemented by developers to specify how to update data stores based on the received logs. On the other hand, to adapt to the execution engines, the adaptor should also provide the \texttt{getMeta()} and \texttt{getData()} interfaces for execution engines to access the necessary metadata and the required data record identified by a given \texttt{key}. The adaptor is correspondingly designed for different execution engines and storage engines. Take a graph query engine and graph-native storage as an example, the adaptor provides a graph data query interface and graph data update interface. Each storage node has also a memory cache for the frequently accessed data to avoid expensive disk access.

\subsection{Conflict Handling in Transaction Layer}
\label{sec:arch:taas}

\Paragraph{Requirements} 
The core of concurrency control (CC) is conflict handling. Considering the requirements of independent transaction service, the conflict handling algorithm used in \taas should satisfy a set of specific requirements. First, the conflict handling should follow \textit{multi-master} architecture, \ie any node can accept updates and works in a peer-to-peer manner. Since the readsets/writesets are naturally sent from different execution engine instances, only allowing single-write would incur a single node bottleneck. In addition, transaction service should be independently scaled, \ie any node can be shutdown or a new node can join at any time, a peer-to-peer decentralized CC algorithm is more reasonable. Second, the conflict handling algorithm should be \textit{optimistic}. Due to the lazy update of the data in the storage layer, the execution layer could read stale data, and a transaction is optimistically executed in the execution layer. To align the needs of optimistic execution, the conflict handling in the transaction layer should also be optimistic. Third, to improve the efficiency of conflict handling, the writes of transactions are usually batched and exchanged with other \taas nodes in \textit{batches}. 

\Paragraph{Epoch-based Multi-Master OCC}
Regarding the conflict handling algorithm in the \taas layer, there could be many choices that satisfy the requirements above. We leverage the epoch-based multi-master OCC \cite{zhou2023geogauss} as the default conflict handling algorithm. On each \taas node, the readset and writeset received from an arbitrary execution node are first tagged with a timestamp and an epoch number according to its local clock. The epoch number increments every few milliseconds (\eg 10 ms by default). Different from FoundationDB \cite{zhou2021foundationdb} and TiDB \cite{huang2020tidb} that use a global timestamp ordering service, the timestamp in \taas is generated by the local server to avoid the single server bottleneck. The readsets and writesets cached by each \taas node are exchanged with every other \taas node at the end of epoch. After receiving the readsets/writesets of the same epoch from all the other peer nodes, each \taas node merges these writesets in terms of a deterministic rule (\eg first-writer-win). Because all the \taas nodes will eventually receive the readsets/writesets of an epoch from all the peer nodes (\ie the same input) and they follow the same merging rule, the conflict merging outputs of an epoch are the same. That is, the merge result is replicated on all \taas nodes, improving the durability. The merge result that indicates the successful updates are logged on local disk, which will be lazily sent to the storage layer. To avoid the synchronization between two consecutive epochs, the exchange of writes is optimistically triggered as long as the local epoch ends, no matter whether the merge of the previous epoch's writes finishes or not. This optimistic exchange would lead to an increased abort rate but improve the throughput and decrease the latency. For more details, please refer to our previous work \cite{zhou2023geogauss}.

\subsection{Implementation}
\label{sec:arch:impl}

We discuss several implementation details of the \taas architecture, including isolation, data consistency, durability, and fault recovery.

\Paragraph{Isolation}
The epoch-based conflict resolution mechanism inherently generates a series of snapshots, so \taas can support \textit{snapshot isolation} by default. In addition, we can perform readset validation to detect whether a previous read is the same as a later read (\ie violating \textit{repeated read isolation}) or cancel readset validation to support \textit{read committed isolation}. Furthermore, \taas can support \textit{serializable isolation} (Serializable Snapshot Isolation \cite{fekete2005making}) by analyzing the dependency graph built upon the readsets/writesets.

\Paragraph{Read Consistency}
Since the logs in the \taas layer are asynchronously pushed to the storage layer, the execution layer might read the stale data, which results in a read consistency problem. We address this problem by associating a \textit{version number} of the storage data and checking whether the read data is the most recent one according to the latest commit version in the \taas layer. If not, it means that the read data is stale, and the transaction will be aborted. Note that, for some applications where the stale read is permitted, these transactions can pass the readset validation phase for improving throughput.

\Paragraph{Durability}
The transaction's durability is mainly ensured in the \taas layer. Due to our multi-master concurrency control algorithm, multiple replicas of the committed updates are logged on all \taas nodes. During the exchange of writesets, the Raft consensus protocol is used to ensure the writesets are received by most of the peer nodes. Multiple replicas of the committed updates are logged to enhance the durability. The logs are transferred to the storage layer, which also consists of several (cross-region) backup nodes. This further improves the safety of data.

\Paragraph{Fault Recovery}
The execution layer is stateless, so a worker failure will not impact the whole system. The storage layer usually leverages cloud storage, which has backup nodes to ensure data safety. A \taas node in the transaction layer could fail. Since the Raft consensus is used to ensure the successful transferring of writesets, the updates will not be lost. Furthermore, since the states in the \taas layer are checkpointed with a series of replicated snapshots (in the format of epoch logs), any recovery node can restore the recent snapshot by requesting from other peer nodes to continue multi-master transaction processing.
\section{Advantages and Case Studies}
\label{sec:advantages}

This section demonstrates the advantages of \taas by showing several case studies and experimental results. 

\Paragraph{Exeprimental Setup}
All the experiments in this paper are launched on a cluster including three execution nodes, three TaaS nodes, and three storage nodes. Each node (Aliyun ecs.r6e.8clarge instance) is equipped with 32 vCPUs and 64GB DRAM. The network bandwidth between nodes is 10 Gbp/s. We use YCSB \cite{ycsb}, TPC-C \cite{tpcc}, and LDBC-SNB \cite{ldbc} benchmarks for performance evaluation. For YCSB, we use one table with 10 columns and 1,000,000 rows. We evaluate three different variations of YCSB workload: 1) YCSB (normal contention): 95\% reads and 5\% writes.  2) YCSB-HC (high contention): 50\% reads and 50\% write. 3) YCSB-RO (read only): all read queries. All YCSB workloads use the Zipfian access distribution. For TPC-C benchmark, we configure it with 100 warehouses, and we execute a workload of 50\% NewOrder and 50\% Payment for focusing on transaction concurrency conflict handling. For LDBC-SN benchmark (graph queries), we use the LDBC-SNB SF10 data set (10GB, 29,987,835 vertices, 176,623,382 edges). Each execution engine is connected with 32 clients by default. \taas is configured with epoch length 10ms, snapshot isolation, and strong read consistency.

\subsection{Empowering NoSQL DBs with TP Capability}
\label{sec:adv:ap}
NoSQL databases are popular because they store data in flexible schemas that scale easily. Types of NoSQL databases include document databases (\eg MongoDB \cite{mongodb} and CouchDB \cite{couchdb}), key-value stores (\eg HBase \cite{hbase}, LevelDB \cite{leveldb}, and Cassandra \cite{cassandra}), and graph databases (\eg NebulaGraph \cite{nebula} and Neo4j \cite{neo4j}). However, most of these existing NoSQL databases do not provide transactional semantics with ACID properties. Every data operation takes place in a single transaction. Users do not have the ability to group multiple operations into a single transaction. In this subsection, we will show that by connecting existing NoSQL databases to \taas, they can be empowered with TP capability. We only take HBase as an illustrative example, but we have also successfully connected another KVS LevelDB and a graph database NebulaGraph to \taas. The results of the TP performance empowered by \taas are promising.

\begin{figure}[t!]
\vspace{-0.2in}
  \centerline{\includegraphics[width=3.3in]{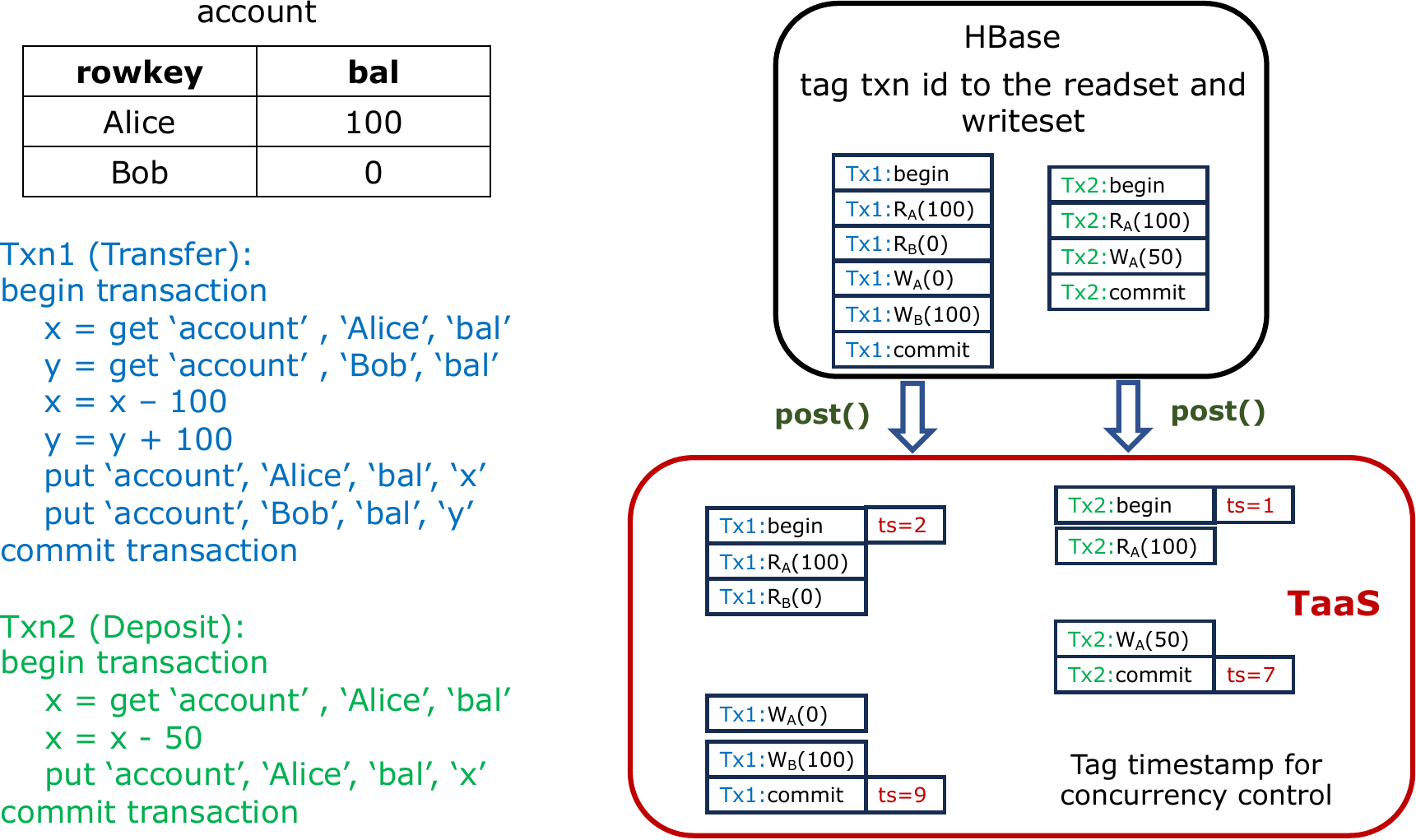}}
  \centering
  \vspace{-0.1in}
  \caption{An illustrative example of HBase connecting to TaaS for supporting ACID transactions.}
  \label{fig:tp}
  \vspace{-0.1in}
\end{figure}

\Paragraph{Empowering HBase/LevelDB/Nebula with TP Capability}
As shown in Figure \ref{fig:tp}, we illustrate how we extend a distributed key-value store HBase to connect \taas and gain the TP capability. Given an \texttt{account} table with \texttt{name} as the rowkey and \texttt{balance} as the value. Users specify a \texttt{Transfer} transaction and a \texttt{Deposit} transaction. These two transactions consist of multiple \texttt{get()} and \texttt{put()} operations. HBase should be slightly modified to tag transaction ids to the readsets and writesets of the corresponding transactions. These readsets and writesets are posted to the \taas layer, where a transaction begin timestamp and a transaction commit timestamp are tagged for each transaction, which are used for conflict handling based on our OCC algorithm (see \S\ref{sec:arch:taas}). The conflict of these two transactions is detected and resolved in the \taas layer, and then the successful writes are logged. We also connect a standalone key-value store LevelDB and a graph database to \taas in the same way to support concurrent transaction processing.

\begin{figure}
\vspace{-0.4in}
		\centering
  		\subfloat[HBase throughput]{\label{fig:ap:hbase}             \includegraphics[width=0.22\textwidth]{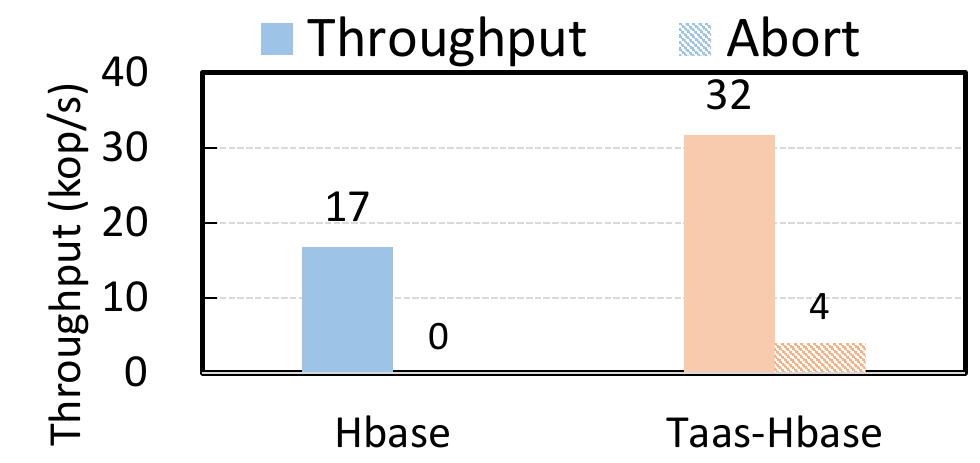}}
    		\subfloat[HBase average latency]{\label{fig:ap:hbase_latency}        \includegraphics[width=0.22\textwidth]{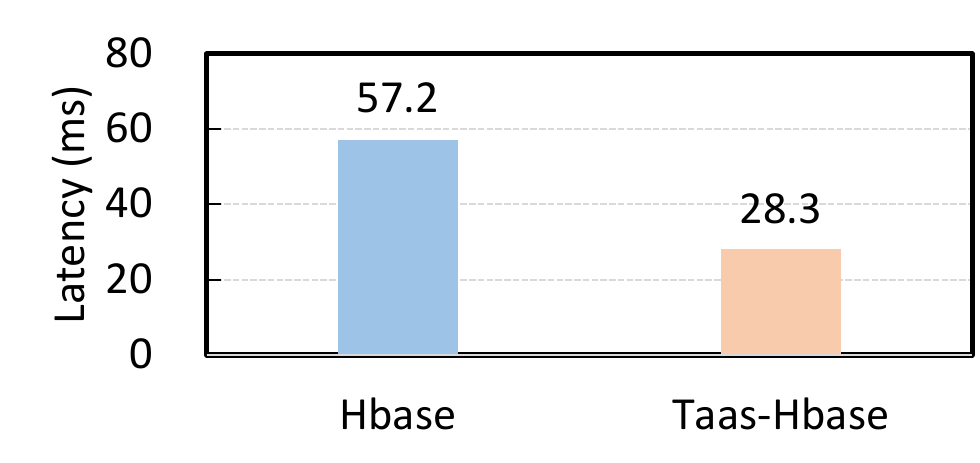}}\\\vspace{-0.1in}
		\subfloat[LevelDB throughput]
            {\label{fig:ap:leveldb}           \includegraphics[width=0.22\textwidth]{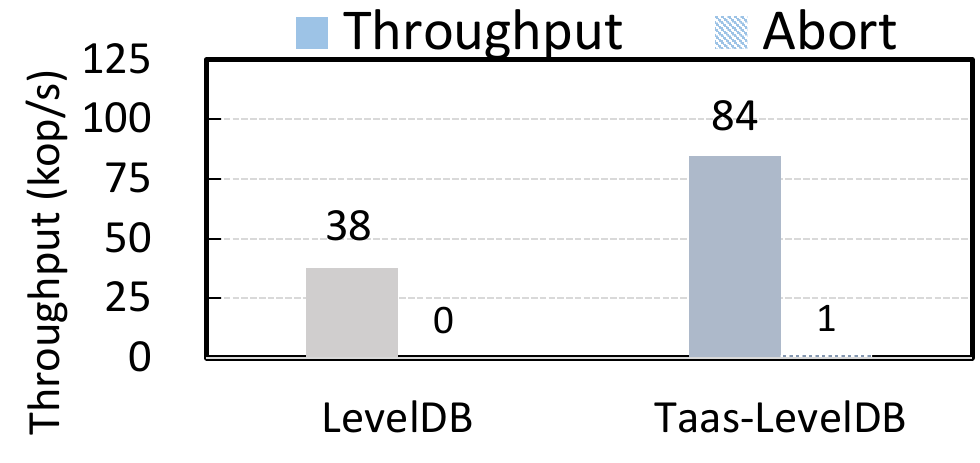}}
          \subfloat[LevelDB average latency]{\label{fig:ap:leveldb_latency}      \includegraphics[width=0.22\textwidth]{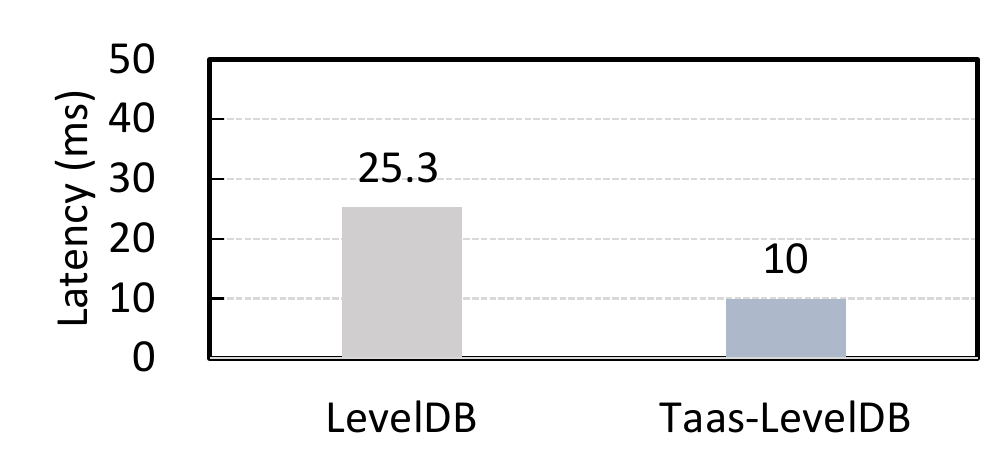}}\\\vspace{-0.1in}
            \subfloat[Nebula throughput]
            {\label{fig:ap:nebula}                     \includegraphics[width=0.22\textwidth]{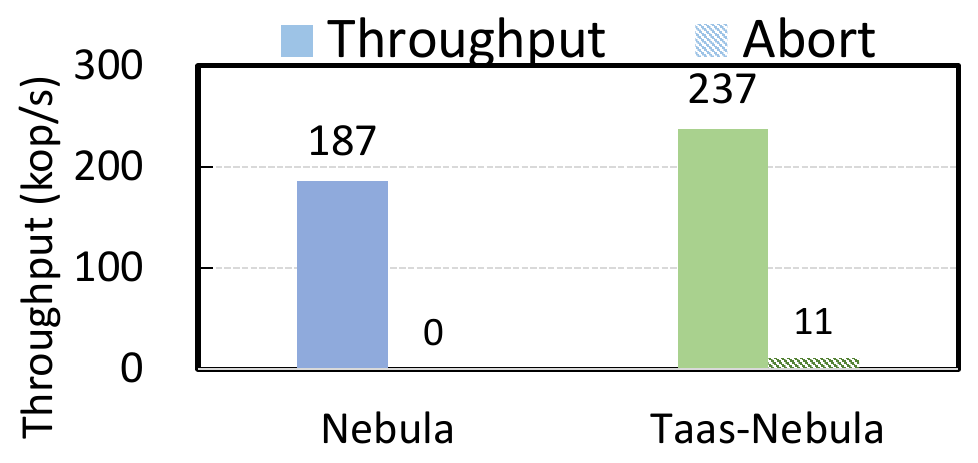}}
            \subfloat[Nebula average latency]{\label{fig:ap:nebula_latency}                \includegraphics[width=0.22\textwidth]{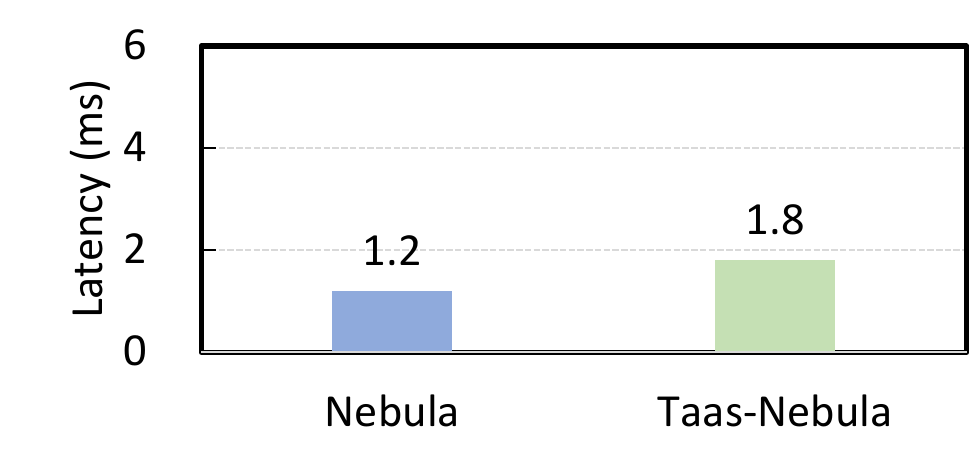}}
            \vspace{-0.1in}
        \caption{TP performance provided by \taas. Throughput: operations per second. Latency: average latency per operation.}
		\label{fig:support-tp}
  \vspace{-0.1in}
\end{figure}
 
\Paragraph{Experimental Results}
To demonstrate the TP performance provided by \taas, we evaluate the operation-based throughput and average latency of original NoSQL databases and that of the \taas-enhanced ones. Note that, the original HBase, LevelDB, and NebulaGraph only support single-operation transactions, so we measure the number of operations per second rather than the number of transactions per second. Regarding the latency, we measure the average latency of each operation rather than that of each transaction. To make YCSB/LDBC-SN a transactional benchmark, we wrap 10 random read/write operations in each transaction. As shown in Figure \ref{fig:support-tp}, for HBase, LevelDB, and NebulaGraph, the throughput is not impacted after connecting \taas. In contrast, by connecting to \taas, these NoSQL databases show higher operation throughput and lower latency due to concurrent execution supported by \taas.

\subsection{Making Standalone TP Engine Distributed}
\label{sec:adv:disttp}

Many open-source standalone databases provide strong and stable TP modules, such as MySQL \cite{mysql}, PostgreSQL \cite{pg}, openGauss \cite{opengauss}, \etc Their communities are active and evolved continuously. However, their original versions do not support distributed TP. It is hard to gain TP's horizontal scalability by relying on them. Fortunately, by connecting multiple standalone TP engine instances to \taas, we can achieve distributed TP easily. We will illustrate how to achieve this goal by connecting multiple openGauss instances to \taas.

\begin{figure}[t!]
\vspace{-0.2in}
  \centerline{\includegraphics[width=3.3in]{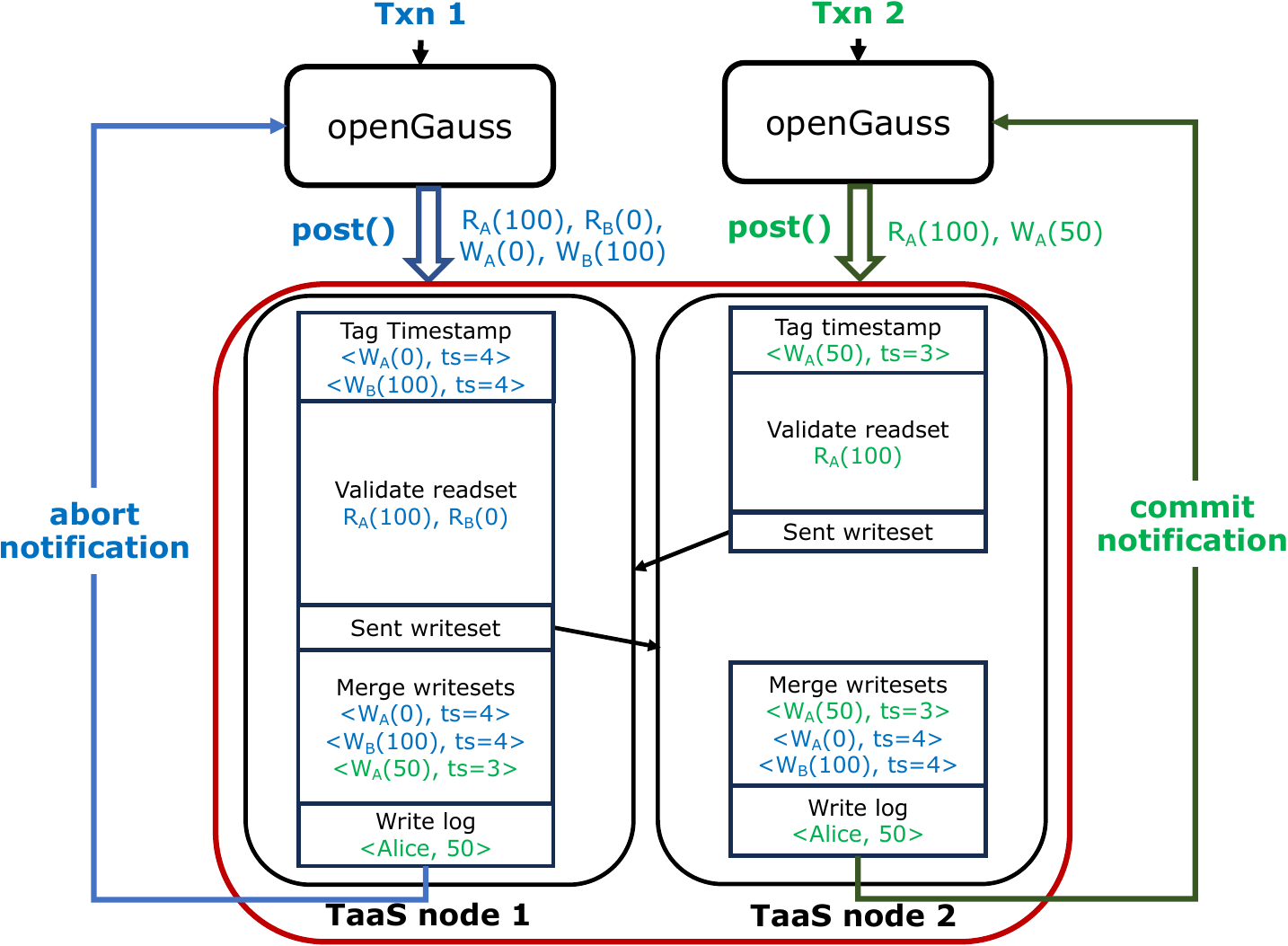}}
  \centering
  \vspace{-0.1in}
  \caption{An illustrative example of multiple openGauss instances connecting to TaaS for supporting multi-master distributed TP and horizontal scalability.}
  \label{fig:dist}
  \vspace{-0.1in}
\end{figure}

\Paragraph{Connecting Multiple openGauss Instances to Support Multi-Master Distributed TP}
We chose openGauss\cite{opengauss}, an open-source standalone relational database system, to connect \taas for illustration. As depicted in Figure \ref{fig:dist}, after execution of a transaction, openGauss is slightly modified to post the readset and writeset to the \taas layer. The readset and writeset of a transaction are randomly sent to a \taas node. Each \taas node tags the writeset with a local timestamp and performs a readset validation to check whether a certain isolation requirement is violated (read-write conflict). For example, the repeated read isolation is violated if the read data version is not the latest, \ie the data version is updated by other transactions when receiving a write of this data item. Suppose a transaction passes the readset validation, the writeset of this transaction is exchanged with other \taas nodes at the end of each epoch. Then a writeset merge operation is performed to check the write-write conflicts. All the \taas nodes follow the same writeset merge rule, so the merge result is the same. For example in Figure \ref{fig:dist}, two concurrent transactions (\ie the two transaction in Figure \ref{fig:support-tp}) both write $A$, \ie one writes $A=0$ and another one writes $A=50$. Following the same first-write-win rule, the write of $A=50$ has a smaller timestamp $ts=3$ than the write of $A=0$, so transaction 2 with $A=50$ is successfully committed while transaction 1 with $A=0$ is aborted. By connecting more \taas nodes, the transaction processing performance can be improved since the readset validation overhead is distributed. But too many \taas nodes will result in heavy network traffic due to exchanging writesets among nodes.

\begin{figure}[t!]
\vspace{-0.1in}
  \centerline{\includegraphics[width=2.3in]{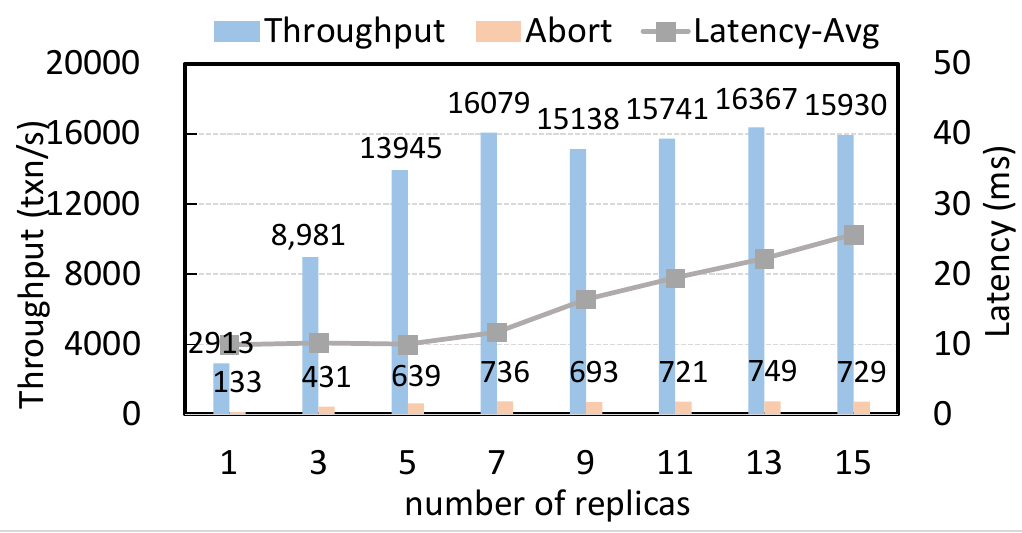}}
  \centering
  \vspace{-0.1in}
  \caption{Scaling performance of distributed TP.}
  \label{fig:scalability}
  \vspace{-0.1in}
\end{figure}

\Paragraph{Experimental Results}
To demonstrate the TP's horizontal scalability by connecting to \taas, we conduct an experiment with varying \taas nodes. By adding more \taas nodes, the workload is distributed to more conflict handling nodes, for improving the TP performance. We vary the number of \taas nodes from 1 to 15. The throughput and latency results are shown in Figure \ref{fig:scalability}. The throughput is steadily increased from 1 node to 7 nodes, and the latency is stable around 10ms. However, when using more than 7 nodes, the throughput is not increased, and the latency is increasing. This implies that the network costs outweigh the benefits of distributed TP. Leveraging low-latency and high-throughput communication protocols, such as Remote Direct Memory Access (RDMA) and Compute Express Link (CXL)), is expected to greatly improve the performance. Nevertheless, we can see that connecting multiple standalone TP engines to \taas can achieve horizontal scalability.

\subsection{Supporting Multi-Model Transactions}
\label{sec:adv:multimodel}

Databases use a variety of data models, with document, graph, and key–value models being popular. A multi-model database is a database that can store, index, and query data in more than one model. However, there are two major challenges to building such a multi-model database. First, it leads to a significant increase in operational complexity. Second, there is no support for maintaining data consistency across the separate data stores. Fortunately, \taas delivers a trivial solution to tackle these two challenges. 

\begin{figure}[t!]
\vspace{-0.2in}
  \centerline{\includegraphics[width=3.4in]{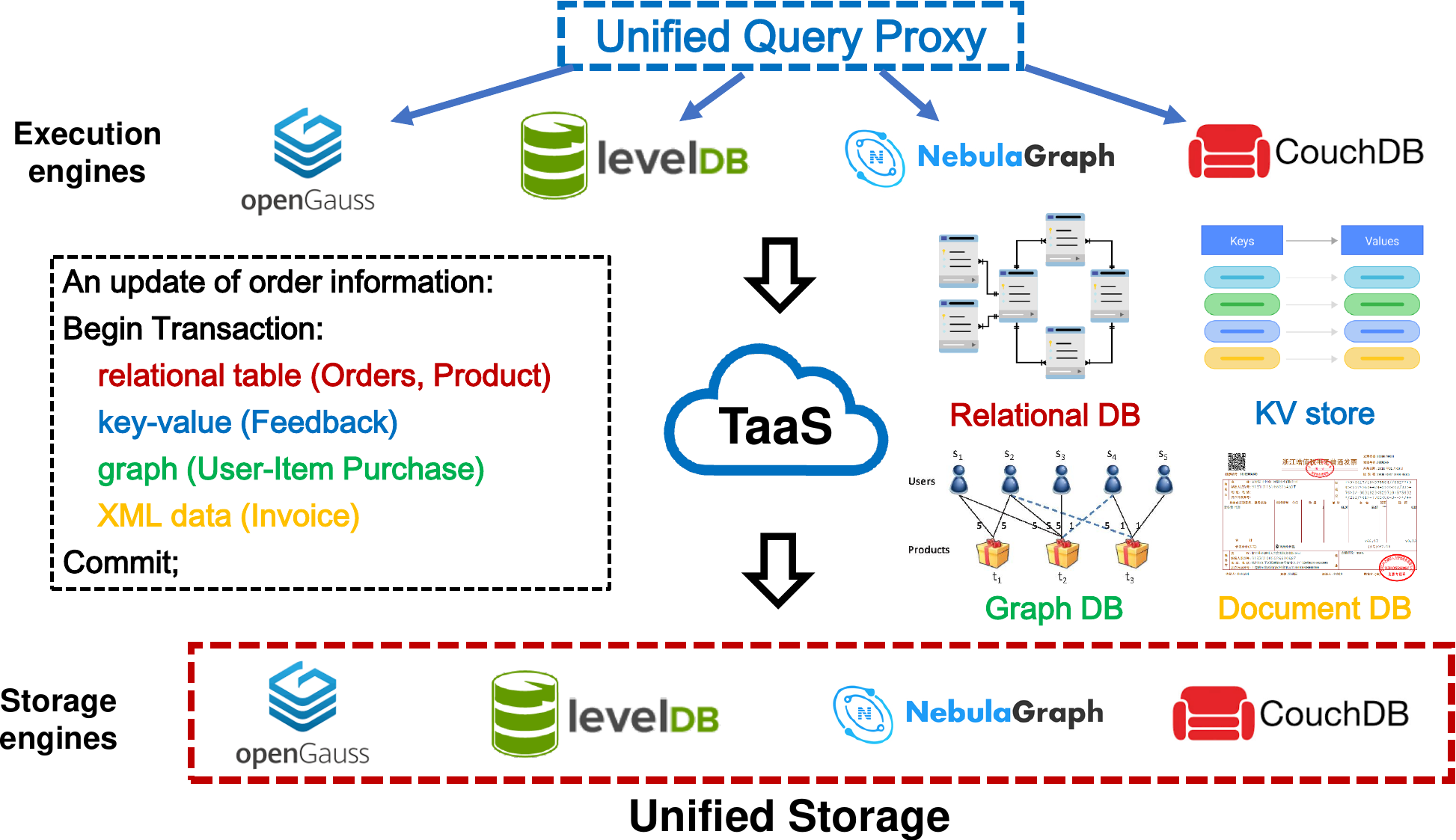}}
  \centering
  \vspace{-0.1in}
  \caption{Multi-model transaction support by unifying multiple execution engines and storage engines through \taas.}
  \label{fig:multimodel}
  \vspace{-0.1in}
\end{figure}

Figure \ref{fig:multimodel} illustrates how to support multi-model transactions involving relational database openGauss \cite{opengauss}, key-value store LevelDB \cite{opengauss}, graph database NebulaGraph \cite{nebula}, and document database CouchDB \cite{couchdb}. By connecting multiple execution engines with different data models to \taas, we can create a unified query proxy to decompose a multi-model transaction into multiple sub-transactions (each corresponding to a data model) and distribute these sub-transactions to different execution engines. Therefore, we can support multi-model transactions by leveraging existing execution engines with only slight modifications. Furthermore, the \taas layer can be thought of as a data consistency ensurance layer. The data consistency problems across separate data stores are resolved by \taas. Then, we will have a unified storage containing multiple data stores with different models.

\Paragraph{Unifying LevelDB (KV), openGauss (SQL), and Nebula (Graph) to Support Multi-Model TP} 
For illustration of supporting multi-model transactions, we unify LevelDB (KV) \cite{leveldb}, openGauss (SQL) \cite{opengauss}, and Nebula (Graph) \cite{nebula} by connecting them to \taas. The sub-transactions of a multi-model transaction are labeled with the same transaction ID, which will be used in the merging of writesets in the \taas layer.
Each engine executes the corresponding sub-transactions optimistically and posts the resulting readsets/writesets to \taas.
The epoch-based multi-master OCC algorithm (\S\ref{sec:arch:taas}) is used to resolve cross-model read-write or write-write conflicts. It is noticeable that the readsets/writesets of the sub-transactions that belong to the same multi-model transaction should be routed to the same \taas node, which is essential for guaranteeing atomicity, \ie the \taas node should know the commit or abort information of each sub-transaction. 

\begin{figure}
\vspace{-0.2in}
		\centering
		\subfloat[Throughput]{\label{fig:multi}       \includegraphics[width=0.22\textwidth]{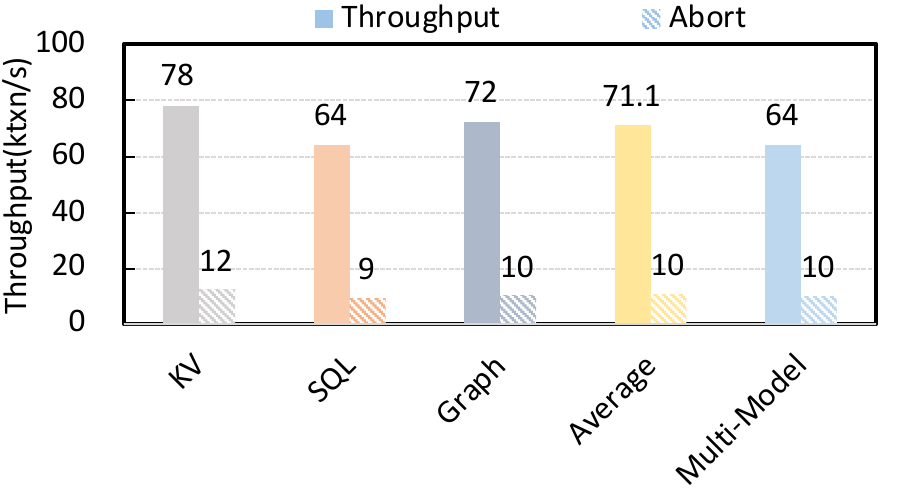}}
		\subfloat[Latency]{\label{fig:multi_latency}  \includegraphics[width=0.22\textwidth]{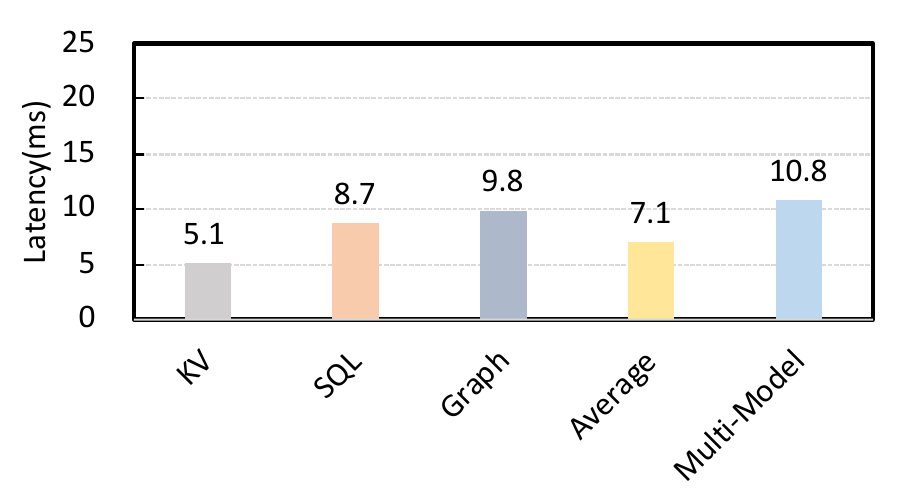}}
		\vspace{-0.15in}
		\caption{Performance of multi-model transactions.}
		\label{fig:modelresults}
  \vspace{-0.1in}
\end{figure}

\Paragraph{Experimental Results}
To verify the performance of our cross-model transactions (by unifying LevelDB, openGauss, and NebulaGraph), we generate a synthetic cross-model transactional
workload with 9 operations per transaction, including 4 KV operations generated by YCSB \cite{ycsb}, 4 SQL operations generated by TPC-C \cite{tpcc}, and 1 graph operation generated by LDBC-SNB \cite{ldbc}). We measure the throughput (number of operations per second) and the latency (average latency per operation) of multi-model transactions. The results are shown in Figure \ref{fig:modelresults}. For comparison, we also generate single-model transaction workload (through YCSB, TPC-C, and LDBC-SN, respectively) to test each single-model engine. We also report the weighted average throughput and latency of these single-model transactions. We can see that the KV engine is the fastest and the SQL engine is the slowest. The performance of multi-model transactions is determined by the slowest engine since the transaction cannot commit until all engines return responses.

\subsection{High Performance}
\label{sec:adv:perf}

\begin{figure}
\vspace{-0.15in}
		\centering
		\subfloat[Throughput]{\label{fig:overall}         \includegraphics[width=0.22\textwidth]{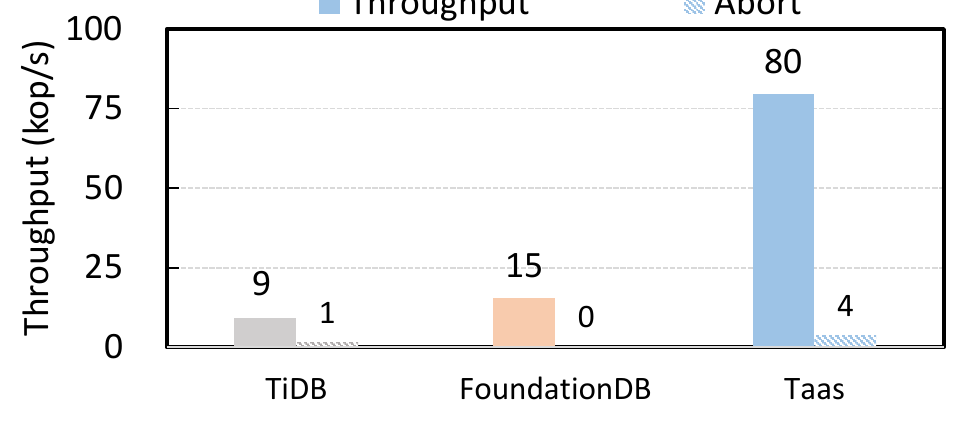}}
		\subfloat[Latency]{\label{fig:overall_latency}    \includegraphics[width=0.22\textwidth]{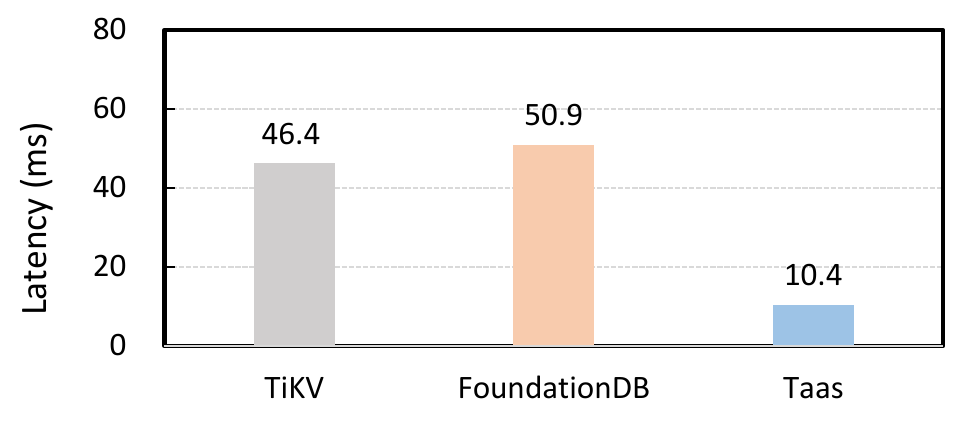}}
		\vspace{-0.15in}
		\caption{Comparison with TiDB and FoundationDB.}
		\label{fig:performance}
  \vspace{-0.15in}
\end{figure}

To demonstrate the performance superiority, we 
run comparison experiments with the decoupled databases TiDB \cite{huang2020tidb} and FoundationDB \cite{zhou2021foundationdb}. The performance comparison results are shown in Figure \ref{fig:performance}. \taas follows the multi-master architecture, exchanging writesets with each other to complete conflict resolving and data backup simultaneously. This helps reduce network communication and transaction latency. However, TiDB and FoundationDB both require an additional Paxos process to backup logs before returning commit notifications to users, which slows down the process.
In addition, \taas leverages asynchronous log push-downs to avoid synchronization overhead, which increases the throughput and reduces transaction delays.

\section{Challenges and Opportunities}
\label{sec:challenges}

\taas shows promising results and potentials in improving cloud-native databases. The key benefit that attracts users using \taas is the powers and functions that the TP service itself can provide. In this section, we list several challenges and research opportunities for improving and using \taas.

\begin{enumerate}[leftmargin=0cm]
    \item \textbf{Transaction Affinity Routing.} In our design, an execution node sends a transaction's readset/writeset to a random \taas node. Actually, it shows the affinity between transactions if their reads/writes are overlapped. Transaction affinity routing will define a routing table that routes a transaction to the \taas node that most likely has related transactions, so that most conflicts can be processed locally rather than being merged across workers. This would greatly reduce the network I/Os.
    \item \textbf{Heterogeneity-Aware Adaptor.} The execution engines and the cluster environments exhibit heterogeneity nowadays. For example, GPU execution engines \cite{he2009relational, wang2023hytgraph, wang2019sep} are becoming popular for processing complex tasks, and RDMA/CXL-enabled clusters are more powerful for distributed processing. We should design new adaptors (including new communication protocols) to leverage these new hardwares.
    \item \textbf{NVM-Native \taas.} Non-volatile memory (NVM) with near DRAM speed, lower power consumption, large memory capacity, and non-volatility in light of power failure, promises signifcant performance potential for TP \cite{liu2021zen, arulraj2016write, ji2023falcon}. However, we need to design record-based data organization and new fault recovery mechanisms to maximize the benefits of NVM. In our \taas architecture, NVM is ideal due to its byte-addressable and non-volatile properties.
    \item \textbf{Rich Isolation and Consistency Choices.} The current \taas supports several isolation levels, \eg read committed, repeated read, snapshot isolation, and serializable snapshot isolation. It is desirable to explore more isolation levels and multi-version support for various application requirements. In addition, the consistency of transactions among \taas nodes and the consistency across the \taas layer and storage layer should also be further studied. Adjustable consistency \cite{li2012making} deserves extensive study since it provides substantial performance gains without sacrificing consistency.   
    \item \textbf{Cross-Region TP and Global Data Consistency Layer.} Achieving efficient cross-region geo-distributed TP is challenging due to the high latency and heterogeneous networks. Existing works, such as Google Spanner \cite{corbett2013spanner} and CockroachDB \cite{taft2020cockroachdb}, offer cross-region TP solutions but are closely coupled with distributed execution engines. If \taas supports cross-region TP, any node in any continent can connect to \taas to solve the data consistency problem across regional servers. We can build a global data consistency layer to connect all nodes worldwide.
    \item \textbf{Cross-Device TP and IoT Data Consistency Layer.} There is a need to share data among our IoT devices, \eg smartphone, tablet, laptop, and smartwatch. We could concurrently make payments, make commodity trading, modify a shared document, and so on. Furthermore, these devices may be managed by different companies. By connecting these IoT devices to \taas as a third-party service, we can build a cross-device data consistency layer ensuring fairness.
\end{enumerate}

\eat{
\begin{acks}
 This work was supported by the [...] Research Fund of [...] (Number [...]). Additional funding was provided by [...] and [...]. We also thank [...] for contributing [...].
\end{acks}
}

\clearpage
\balance
\bibliographystyle{ACM-Reference-Format}
\bibliography{acmart}

\end{document}